\pdfoutput=1
\documentclass{emulateapj}  

\usepackage{graphicx} 

\usepackage{apjfonts} 
\usepackage{amsmath}
\usepackage{natbib}
\usepackage{verbatim}
\usepackage{rotating}
\usepackage{lscape}

\slugcomment{ApJ Letters, accepted 7/9/08}
\shorttitle{The Far 3-Kpc Arm} 
\shortauthors{Dame and Thaddeus }

\newcommand{\etal}{{\rm et al.}}

\newcommand{\kms}{km s$^{-1}$}
\newcommand{\msun}{M$_\odot$}
\newcommand{\rsun}{R$_\odot$}

\begin{document}

\title{A New Spiral Arm of the Galaxy: The Far 3-Kpc Arm} 
\author{T. M.\ Dame and P.\ Thaddeus}
\affil{Harvard-Smithsonian Center for Astrophysics, 60 Garden Street, Cambridge MA 02138  \\ tdame@cfa.harvard.edu, pthaddeus@cfa.harvard.edu}

\begin{abstract} 
We report the detection in CO of the far-side counterpart of the well-known expanding 3-Kpc Arm in the central region of the Galaxy. In a CO longitude-velocity map at $b = 0^{\circ}$ the Far 3-Kpc Arm can be followed over at least $20^{\circ}$ of Galactic longitude as a faint lane at positive velocities running parallel to the Near Arm. The Far Arm crosses $l = 0^{\circ}$ at +56 \kms, quite symmetric with the $-53$ \kms expansion velocity of the Near Arm. In addition to their symmetry in longitude and velocity, we find that the two arms have linewidths ($\sim21$ \kms), linear scale heights ($\sim103$ pc FWHM), and H$_2$ masses per unit length ($\sim4.3$ x $10{^6}$ \msun kpc$^{-1}$) that agree to 26\% or better. Guided by the CO, we have also identified the Far Arm in high-resolution 21 cm data and find, subject to the poorly known CO-to-H$_2$ ratio in these objects, that both arms are predominately molecular by a factor of 3--4.  The detection of these symmetric expanding arms provides strong support for the existence of a bar at the center of our Galaxy and should allow better determination of the bar's physical properties. 
\end{abstract}


\section*{}

  Since its identification in 21 cm emission by \citet{vanWoerden57}, the expanding 3-Kpc Arm has remained at once one of the most obvious Galactic spiral arms and the most puzzling. Its structure in 21 cm and CO longitude-velocity diagrams and its absorption of continuum emission toward the Galactic center demonstrate beyond doubt that the arm lies on the near side of the center and is expanding away from it at a velocity of $-53$ \kms at $l = 0^{\circ}$. The arm's large non-circular motion has been attributed to explosive expulsion of gas from the center (van de Kruit 1971; Sanders \& Prendergast 1974) and was central to one of the earliest arguments for the existence of a bar at the Galactic center (de Vaucouleurs 1964). 

On the basis of a reanalysis of the Columbia-CfA-Chile CO survey of the Milky Way \citep{Dame01} in the vicinity of the Galactic center \citep{Bitran97}, we have found clear evidence on the far side of the Galaxy for the long postulated and long sought counterpart of the expanding 3-Kpc Arm. The Far 3-Kpc Arm displays a clear symmetry with its near-side counterpart in longitude and velocity, and once account is taken of its greater distance, very similar physical characteristics as well (Table 1).  

The Near and Far Arms appear in the CO $l$-v diagram of Figure 1 as two inclined, parallel lanes symmetrical in velocity on either side of the Galactic center.  The Far Arm as expected is weaker than the Near, but it can be readily followed over at least $20^{\circ}$ of longitude, starting at $l = -12^{\circ}$, with an average intensity of $\sim0.5$ K, more than 3 times the instrumental noise. The linear fits indicated by the dashed lines yield expansion velocities toward $l = 0^{\circ}$ of $-53.1$ \kms for the Near Arm and $+56.0$ \kms for the Far Arm, and, within the uncertainties, identical velocity gradients with $l$ (Table 1). 

%FIG 1 POS
\begin{figure*}
\centering
%\epsscale{1.}
\plotone{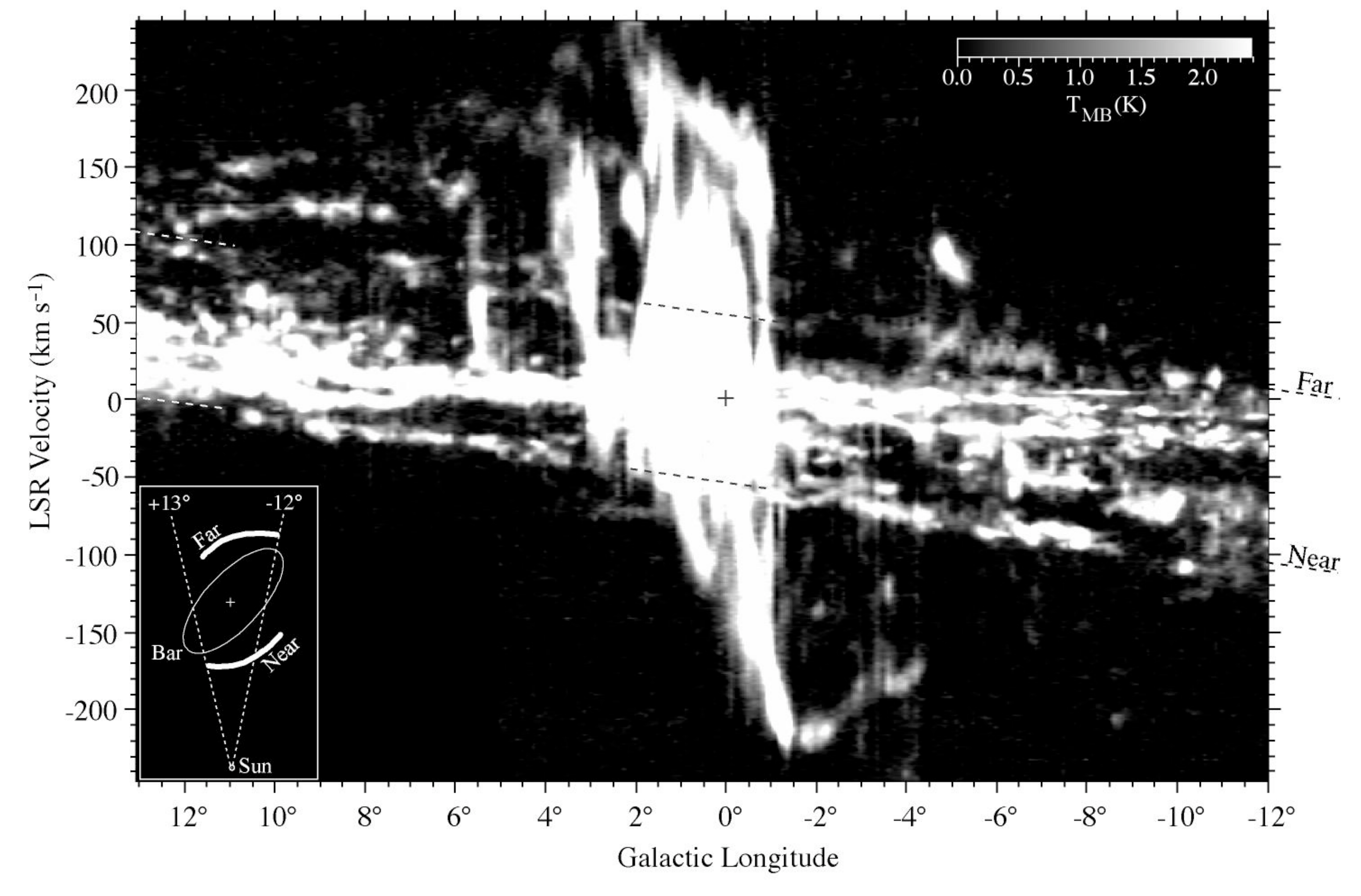}
\caption{Longitude-velocity diagram of CO $1-0$ emission at $b =  0^{\circ}$ from \citet{Bitran97}; this survey is sampled every $7.5\arcmin$ at $|b| \leq 1^{\circ}$ (every $15\arcmin$ elsewhere) with an $8.8\arcmin$ beam to an rms sensitivity of 0.14 K in $1.3$ \kms channels. The dashed lines are linear fits to the inclined, parallel lanes of emission from the Near and Far 3-kpc Arms: for the Near Arm v $= -53.1 + 4.16$ $l$, for the Far v $= +56.0 + 4.08$ $l$. The insert is a schematic showing the approximate locations of the arms over the longitude ranges in which they can be followed clearly in CO. The dotted lines are the limits of the present analysis. 
 }
\label{fig:mapaszoom}
\end{figure*}

To examine the average velocity structure of the Near and Far Arms, smoothing out  fluctuations owing to individual clouds along each arm, we averaged the emission in Figure 1 in narrow linear strips inclined parallel to the arms (which, as noted, are parallel to each other) and labeled each strip by its velocity at $l =0^{\circ}$.  In the resulting plot (Fig. 2), the Near and Far Arms stand out as well-defined peaks to either side of a broad central peak, mainly from the foreground and background disks. Note that the Far Arm is only a factor of two fainter than the Near at $b = 0^{\circ}$, and both are detected at a level far above the instrumental noise. It is also clear from Figure 2 that the CO linewidths of the two arms are similar, with Gaussian fits yielding values of $19.7$ \kms (FWHM) for the Near Arm and $22.2$ \kms for the Far. We expect that further, more refined analyses, which will allow for the expected curvature of the arms with longitude, will result in linewidths that are at most 10\% lower than those we find here. In spite of the difficulty of estimating the overall noise level in Figure 2, which is dominated by the clutter of unrelated emission over which we average, the figure forcefully shows that the detection of the Far 3-Kpc Arm is not marginal, and a similar exercise carried out with  {\it any} moderate-resolution molecular-line survey of the Galactic center should reveal the peak of the Far Arm. 

%FIG 2 POS
%figure
\begin{figure}
\centering
%\epsscale{1.}
\plotone{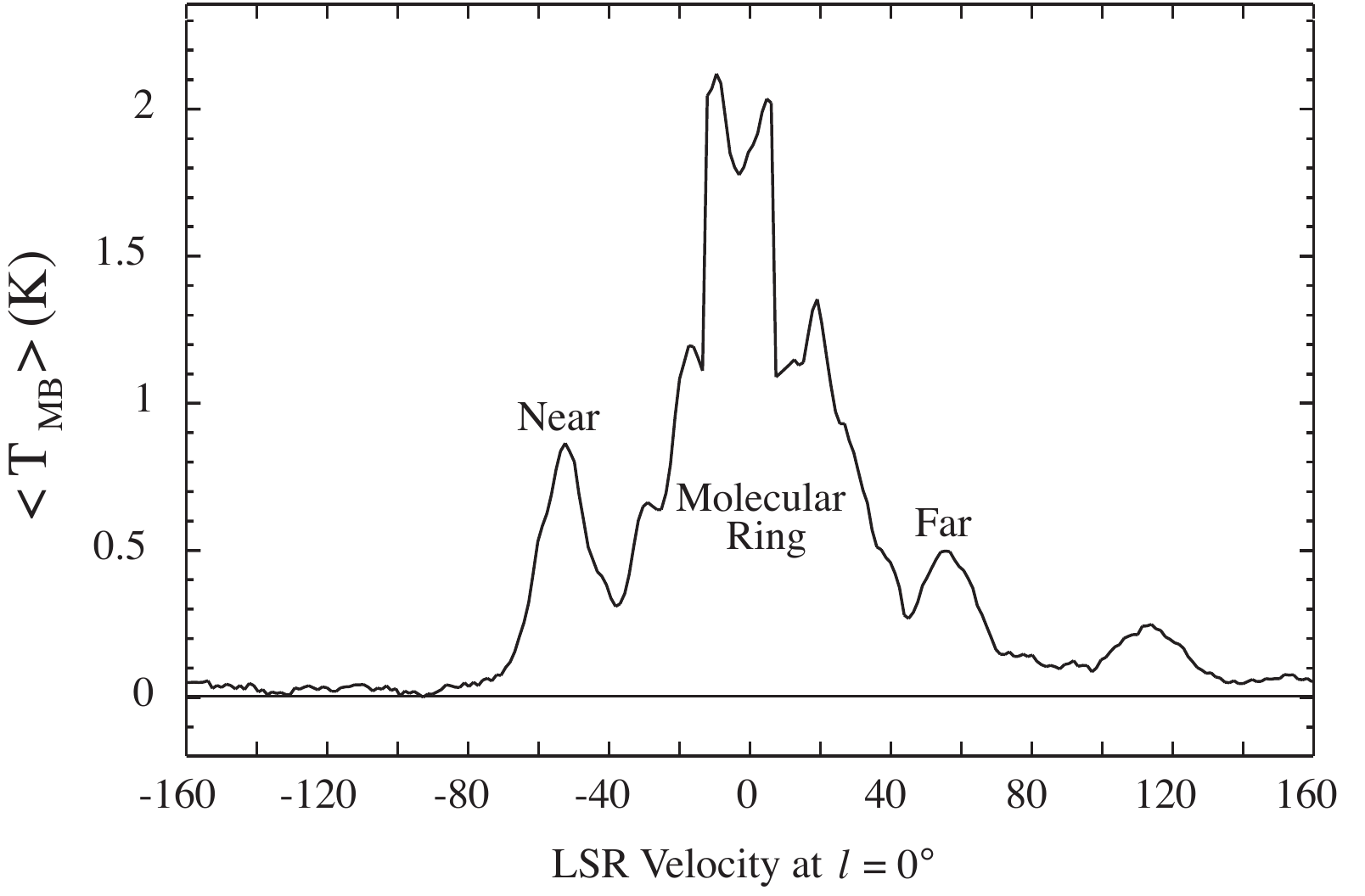}
\caption{
A composite CO spectrum for the Near and Far Arms obtained by averaging the emission in Fig. 1 in narrow linear strips running parallel to the arms, their essentially identical slopes with longitude determined by the linear fits in Fig. 1. The blended regions indicated in Fig. 3 were excluded from the averaging. }
\label{fig:mapas}
\end{figure}

%FIG 3 POS
\begin{figure*}
\centering
%\epsscale{1.}
\plotone{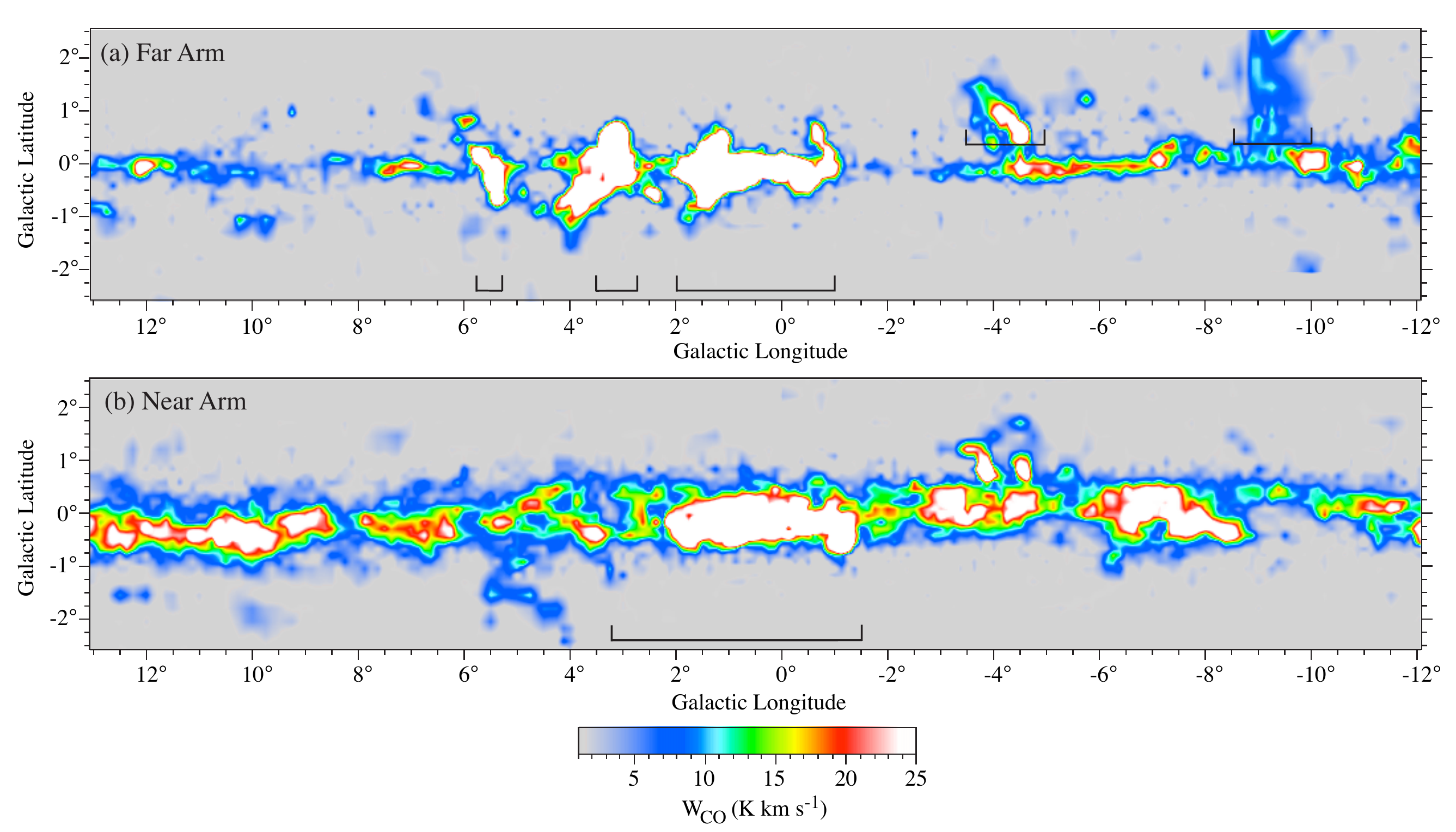}
\caption{The velocity-integrated CO intensity of the Near and Far 3-Kpc Arms. Emission was integrated over a 26 \kms bin centered at each longitude on the arm velocity given by the linear fits in Fig. 1. Brackets indicate blended and other regions excluded in computing the physical properties of the two arms (see Table 1 and the captions to other figures). The color palette shown below is used in both maps. 
 }
 \label{fig:mapaszoom}
\end{figure*}

If the inclined, linear emission feature at positive velocities in Figure 1 is, in fact, on the far side of the Galactic center, it should appear systematically thinner on the sky than does the Near Arm, and in the spatial maps in Figure 3 this is indeed the case. Excluding the blended regions indicated, the angular thicknesses of the two arms appear to be roughly constant along their lengths and to differ by about a factor of two. This result is quantified by the latitude profiles in Figure 4, which yield FWHM thicknesses of $1.1^{\circ}$ for the Near Arm and $0.52^{\circ}$ for the Far. 

Because the Near 3-kpc Arm was first discovered in HI, on finding its far CO counterpart we naturally attempted to identify it in existing 21 cm data. The early 21 cm surveys were too poor in angular resolution and sensitivity to distinguish the thin, faint lane of the Far Arm, but in the recent Australia Telescope (ATNF) 21 cm survey at a resolution of $2\arcmin$ \citep{Griffiths05} there is clear evidence of the Far Arm.  In their Figure 6, a 21 cm $l$-v diagram at $b = 0^{\circ}$, the Far Arm is seen extending from the right  at v~$\sim75$ \kms. At $l < 6^{\circ}$ this arm shows a curious velocity bifurcation that is also seen in the Far CO Arm in this direction and in some segments of the Near CO Arm (e.g., $l = -4^{\circ}$ to $-1^{\circ}$ in Fig. 1). At negative longitudes the Far Arm in HI is largely blended with emission from both distant gas beyond the solar circle and foreground gas in the inner disk; however, a segment of the Far Arm can be traced in the ATNF survey at $l > 354^{\circ}$. 

As we required of the Far CO Arm in Figure 1, the Far HI Arm evident in Figure 6 of McClure-Griffiths et al. should be narrow in latitude, and the HI $b$-v map in Figure 5a shows that indeed  it is. Comparison of this map to the CO $b$-v map in Figure 5b shows that the Far Arm has about the same thickness in both species (see also Table 1).  In the positive longitude range included in Figure 5a, it is the Near Arm that is masked in HI by both distant gas beyond the solar circle and foreground gas in the inner disk.  In contrast, the Near Arm is seen clearly in CO, because there is little CO beyond the solar circle and because the cloud-cloud velocity dispersion of the CO is lower than that of the HI.

%FIG 4 POS
%figure
\begin{figure}
\centering
%\epsscale{1.}
\plotone{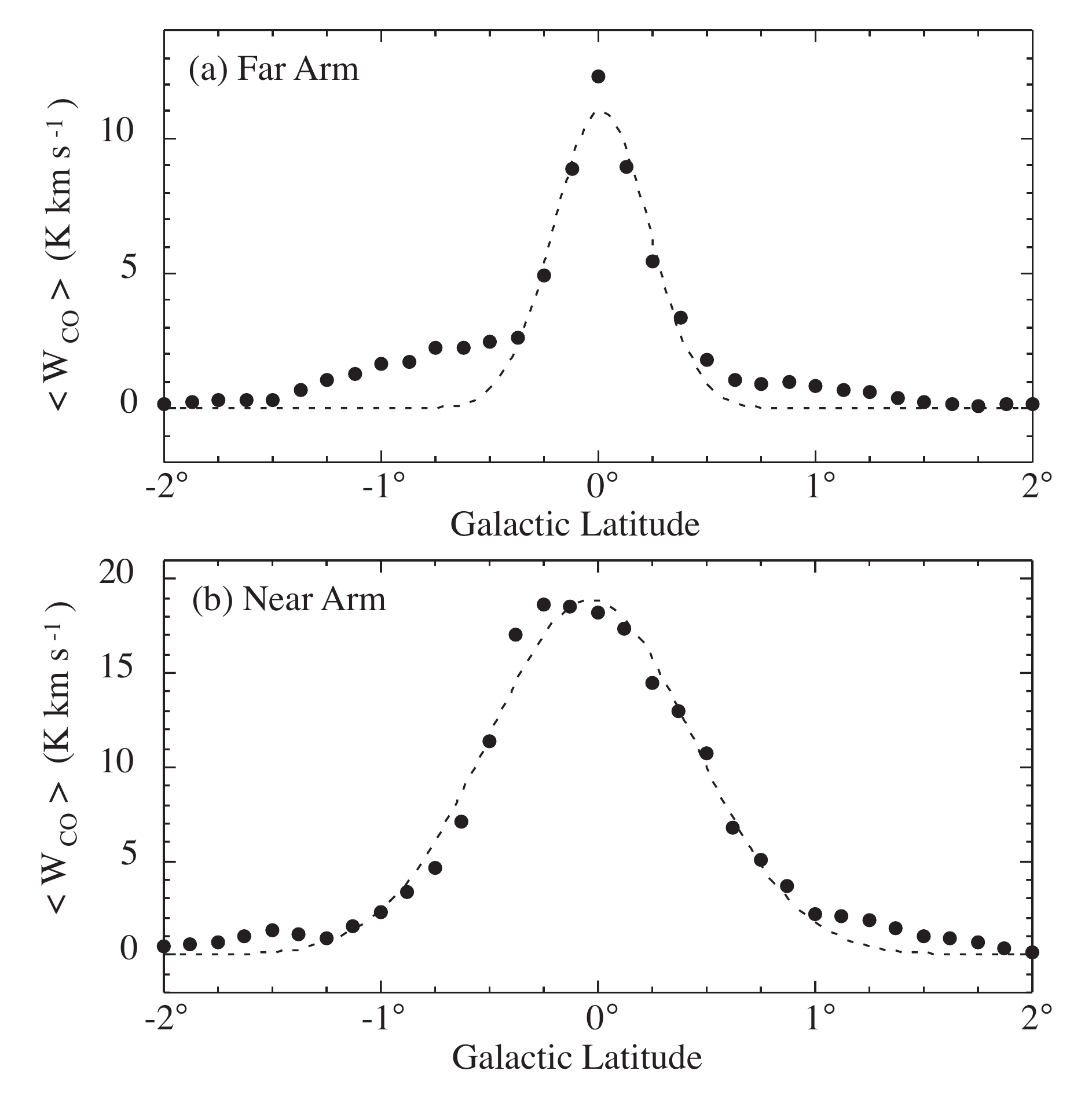}
\caption{Latitude profiles of the CO emission from (a) the Far and (b) Near 3-Kpc Arms, obtained by averaging the spatial maps in Fig. 3 over longitude. The blended regions and two vertical extensions marked in Fig. 3 were excluded. The dotted curves are Gaussian fits. Since the profile of the Far Arm shows high-latitude wings presumably from unrelated foreground material, the fit for this arm was confined to points at $|b| < 0.5^{\circ}$.  
 }
\label{fig:mapas}
\end{figure}

To further quantify the properties of the two arms, distances are required. The Near 3-Kpc Arm derives its name from its Galactic radius as estimated by van Woerden et al. (1957) on the basis of an apparent southern tangent at $l = -22^{\circ}$ and the \rsun value of 8.2 kpc then adopted. Subsequent work by Bania (1980), Cohen \& Davies (1976), and others similarly suggested a possible northern tangent near $l = +23.5^{\circ}$. Although the locations and even the existence of both tangents are still in doubt, here we adopt tangent directions of $\pm23^{\circ}$ for the Near Arm on the basis of careful study of our composite CO survey (Dame et al. 2001).  With \rsun $= 8.5$ kpc, these tangents imply a radius of 3.3 kpc for the Near Arm, and given its similarity in so many other respects, we assume the same radius for the Far Arm. Although the inclinations of the arms are unknown, here we assume for simplicity that the arm distances vary little along their lengths; thus the Near Arm is at a distance of 5.2 kpc (\rsun $-$ R$_{arm}$) and the Far at 11.8 kpc (\rsun + R$_{arm}$).  

%FIG 5 POS
%figure
\begin{figure}
\centering 
%\epsscale{1.}
\plotone{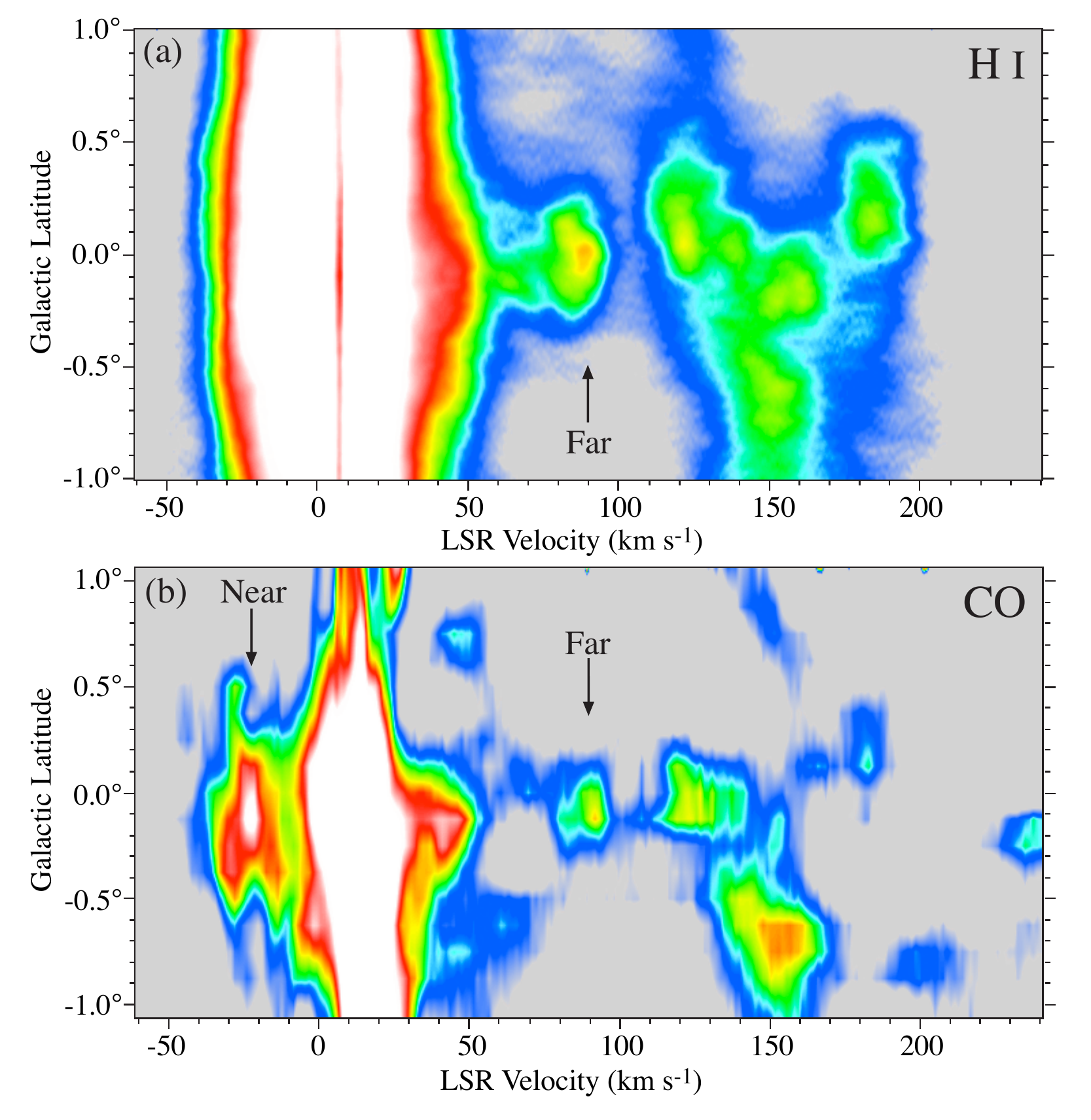}   
\caption{(a) HI latitude-velocity map integrated in $l$ from $5^{\circ}$ to $9^{\circ}$, a range in which the Far Arm is well separated in velocity from foreground 21 cm emission. Data are from McClure-Griffiths et al. (2005).  Colors represent log intensity (K-arcdeg) from gray (1.0) to white (2.5).  
(b) CO latitude-velocity map integrated in $l$ from $2^{\circ}$ to $8^{\circ}$, a range in which both the Near and Far Arms are well defined and which overlaps that of the HI map in (a); two of the blended regions indicated in Fig. 3 were excluded from the integration. Colors represent log intensity (K-arcdeg) from gray (0.0) to white (0.8). 
 }
\label{fig:fig4}
\end{figure}

With these distances, H$_2$ and HI masses for the arms can be computed directly from their integrated luminosities in CO and 21 cm. Although little is known about the value of the CO-to-H$_2$ mass conversion factor in these unusual arms, it is reasonable to suppose that a similar value applies to both. Here we assume a fairly standard Galaxy-wide average value of N(H$_{2}$)/W$_{CO}$ = 1.8 x 10$^{20}$ cm$^{-2}$ K$^{-1}$ km$^{-1}$ s \citep{Dame01}. Excluding the regions of blending indicated in Figure 3, the longitude range considered here covers a 1.47-kpc length of the Near Arm and a 3.24-kpc length of the Far. Over these ranges we find that the two arms have H$_2$ masses per unit length that differ by only 26\% (Table 1).  HI masses were similarly computed from the ATNF 21 cm survey using a well-defined $4^{\circ}$ segment of the Far Arm at positive longitudes and a comparable $9^{\circ}$ segment of the Near Arm at negative longitudes (Table 1). We find that both arms are primarily molecular, the H$_2$ mass exceeding the HI mass in the Near Arm by a factor of 3 and in the Far Arm by a factor of 4 (a conclusion of course sensitive to the adopted CO-to-H$_2$ ratio).

It is worth noting that the Near and Far Arms are found to have almost identical linear thicknesses in H$_2$ and similar thicknesses in HI (Table 1). We speculate that the similarity of the thickness of the HI and CO in the two 3-Kpc Arms, versus the roughly 2:1 ratio found elsewhere in the Galaxy, is the result of the evident absence of star formation in these arms (Lockman 1980), activity that one expects would heat the gas and inflate the HI thickness.  

The idea that the Milky Way is a barred spiral galaxy can be traced back at least fifty years to \citet{Johnson57}, \citet{deV64}, and others, and there is now fairly wide consensus that we are indeed located in a barred spiral \citep{Binney91, Blitz91, Benjamin05}. \citet{Binney08} in their text {\em Galactic Dynamics} state unequivocally that "Our Galaxy is the nearest barred spiral". The delineation here of the predicted two symmetric, expanding 3-Kpc Arms confirms beyond a reasonable doubt that we are located in such a galaxy; further study of these arms should permit better determination of the orientation, size, and other properties of the bar.  Had it been possible to detect the Far Arm when \citet{vanWoerden57} discovered the Near Arm, it seems likely that attempts over 50 years to map the structure of the inner Galaxy would have proceeded more rapidly and directly.  

Because the properties of the Far Arm are so close to those expected for a far-side counterpart of the expanding 3-Kpc Arm, it is natural to ask how the arm escaped notice for so long. Hints of the Far Arm are evident in even the first large-scale CO survey of the region (Bania 1977; see Fig. 1), but not at a level that would have allowed a convincing case to be made. On the contrary, Oort (1977) cited Bania's survey as evidence "That there is no counterpart of the 3-kpc arm behind the center". This view hardened over decades as the so-called "+135 \kms feature" was widely adopted as the only possible far-side counterpart of the expanding arm. The 21 cm surveys with angular resolution adequate to resolve the thin Far Arm have become available only in the past few years (ATNF: McClure-Griffiths et al. 2005, VLA: Stil et al. 2006), and such data for the region within $5^{\circ}$ of the center have yet to be published. Among the factors that masked the Far Arm in molecular line surveys is the common practice of tracing spiral arms with $l$-v maps integrated over latitude (e.g., Fig. 3 of Dame et al. 2001), which leaves the Far Arm unresolved and confused with foreground emission, and the widespread use of contour maps in the past and color maps at present---both of which can sometimes mask weak, large-scale features. The two parallel lanes obvious in Figure 1 are less apparent in 21 cm data even at high angular resolution, because at negative longitudes the Far Arm is badly blended with emission from the outer and inner disks, and at positive longitudes the Near Arm is similarly blended.  

Detection of the Far 3-Kpc Arm immediately suggests many avenues for follow-up study with existing data, new observations, and theoretical studies. Tracking both arms into the more confused regions outside the longitude range considered here---perhaps to their origins at either end of the bar---is a high priority which we are pursuing. The Far Arm should be detectable in other existing spectral line surveys with adequate angular resolution; even those with relatively low sensitivity should reveal the arm in the manner of Figure 2. CO observations with higher sensitivity and angular resolution are needed to better define the Far Arm, in particular, its angular scale height as a function of longitude, which could help constrain the arm's inclination to our line of sight.  Since the Near Arm is known to be deficient in star formation, searches for star formation in the Far Arm are of interest. In addition, delineation of the Far Arm will provide a badly needed new constraint for the hydrodynamical models (e.g., Fux 1999; Bissantz, Englmaier, \& Gerhard 2003) that seek to understand the overall properties of the central bar and its influence on the gas. 

In contrast to the controversy that has long characterized attempts to determine the structure of the inner Galaxy, the 3-Kpc Arms together stand out clearly as an unambiguous, beautifully symmetric structure. \citet{Rougoor60}, in one of their first papers on the expanding 3-Kpc Arm, noted that "The arm is very well defined, and more homogeneous in density as well as velocity than any of the outer arms." That statement is even more true of the twofold larger structure that we can now trace in the inner Galaxy.
 
%________________________________________________________________

\acknowledgments 
We are indebted to T. Bania, R. Benjamin, J. Binney, A. Toomre, and S. Tremaine for highly informative discussions and N. McClure-Griffiths for providing the ATNF 21 cm survey. 

%
%________________________________________________________________

%
%Bibliography
%

\begin{table*}
\footnotesize
\begin{center}
\caption{Comparison of Near and Far 3-Kpc Arms\label{tbl-1}}
\begin{tabular}{ccccccccc}
\tableline\tableline
Arm & d & v$_0$\tablenotemark{a} & dv$/dl$\tablenotemark{a} & $\Delta$v\tablenotemark{b} & $H_2$ dM/d$l$\tablenotemark{c} & HI dM/d$l$\tablenotemark{d} & $\Delta$z(H$_2$)\tablenotemark{e} & $\Delta$z(HI)\tablenotemark{d,e} \\

 & (kpc) & (\kms) & (\kms deg$^{-1}$) & (\kms) & ($10{^6}$ \msun kpc$^{-1}$) &  ($10{^6}$ \msun kpc$^{-1}$) & (pc) & (pc) \\

\tableline
Near & 5.2   & --53.1 & 4.16 & 19.7   & 4.8  & 1.6 & 101 & 138 \\
Far   & 11.8 & +56.0 & 4.08 & 22.2  & 3.8  & 0.9 & 105 & 109 \\
\tableline
\end{tabular}
\end{center}
% Any table notes must follow the \end{tabular} command.
\tablenotetext{a}{Velocity at $l = 0^{\circ}$ and velocity gradient with longitude from linear fits to Fig. 1}

\tablenotetext{b}{Mean velocity width (FWHM) from Gaussian fits to composite profiles in Fig. 2}

\tablenotetext{c}{Averaged $l = -12^{\circ}$ to $8^{\circ}$, excluding blended regions:  $l = -1.5^{\circ}$ to $3.25^{\circ}$ for Near Arm, $-1^{\circ}$ to $2^{\circ}$, $2.75^{\circ}$ to $3.5^{\circ}$, and $5.25^{\circ}$ to $5.75^{\circ}$ for the Far Arm (see Fig. 1)}

\tablenotetext{d}{Averaged over the arm segments best defined in HI:  $l = -9.5^{\circ}$ to $-1.5^{\circ}$ for the Near Arm and $l =5^{\circ}$ to $9^{\circ}$ for the Far}

\tablenotetext{e}{Vertical linear thickness (FWHM); see Fig. 4 for H$_2$;  HI profiles not shown}

\end{table*}

\end{document}